\documentclass[letterpaper,american,pre,aps,superscriptaddress]{revtex4}
\usepackage[latin1]{inputenc}
\usepackage{bm}
\usepackage{multirow,amssymb,amsbsy,amsmath}
\usepackage{stmaryrd}
\usepackage{graphicx}
\usepackage{hyperref}
\makeatletter
\usepackage{pifont}
\makeatother

\begin{document}

\title{Non-Markovian quantum dynamics and the method of correlated projection superoperators}

\author{Heinz-Peter Breuer}

\email{breuer@physik.uni-freiburg.de}

\affiliation{Physikalisches Institut, Universit\"at Freiburg,
             Hermann-Herder-Strasse 3, D-79104 Freiburg, Germany}

\date{\today}

\begin{abstract}
Efficient methods for the description of the non-Markovian
dynamics of open systems play an important role in many proposed
applications of quantum mechanics. Here we review some of the most
important tools that are based on the projection operator
techniques of nonequilibrium statistical mechanics. The standard
product-state projection is generalized to a new class of
correlated projection superoperators that allow the treatment of
strong memory effects, and lead to a non-Markovian generalization
of the Lindblad equation.
\end{abstract}

\maketitle

\section{Introduction}
Relaxation and decoherence processes are key features of the
dynamics of open quantum systems \cite{TheWork}. In the standard
approach one tries to develop appropriate master equations for the
open system's reduced density matrix $\rho_S$ which is given by
the partial trace taken over the environmental variables coupled
to the open system. Invoking the weak-coupling assumption one can
formulate in many cases of physical interest a Markovian quantum
master equation for the reduced density matrix, expressing the
dynamical laws for the irreversible motion of the open system.

However, the theoretical description of quantum mechanical
relaxation and decoherence processes often leads to a
non-Markovian dynamics which is determined by pronounced memory
effects. Strong system-environment couplings \cite{REIBOLD,HU},
correlations and entanglement in the initial state
\cite{INGOLD,BUZEK}, interactions with environments at low
temperatures and with spin baths \cite{LOSS}, finite reservoirs
\cite{GEMMER2005,GMM}, and transport processes in nano-structures
\cite{FOURIER} can lead to long memory times and to a failure of
the Markovian approximation.

Here, we will review the most important features of a systematic
approach to non-Markovian quantum dynamics which is known as
projection operator technique \cite{NAKAJIMA,ZWANZIG,HAAKE,KUBO}.
This technique is based on the introduction of a certain
projection superoperator ${\mathcal{P}}$ which acts on the states
of the total system. The superoperator ${\mathcal{P}}$ expresses
in a formal mathematical way the idea of the elimination of
degrees of freedom from the complete description of the states of
the total system. Namely, if $\rho$ is the full density matrix of
the composite system, the projection ${\mathcal{P}}\rho$ serves to
represent a certain approximation of $\rho$ which leads to a
simplified effective description of the dynamics through a reduced
set of relevant variables.

With the help of the projection operator techniques one derives
closed dynamic equations for the relevant variables
${\mathcal{P}}\rho$. We will discuss two different approximation
schemes. The first one is based on the Nakajima-Zwanzig equation
\cite{NAKAJIMA,ZWANZIG} which represents an integrodifferential
equation for ${\mathcal{P}}\rho$ with a certain memory kernel. The
second scheme employs a time-convolutionless master equation for
${\mathcal{P}}\rho$, i.e a time-local differential equation with a
time-dependent generator
\cite{SHIBATA77,SHIBATA79,SHIBATA80,SHIBATA99,ROYER2003,BREUER2004}.
These equations are used as starting point for the derivation of
effective master equations through a systematic perturbation
expansion.

In the standard approach to the dynamics of open systems one
chooses a projection superoperator which is defined by the
expression ${\mathcal{P}}\rho = \rho_S \otimes \rho_0$, where
$\rho_0$ is some fixed environmental state. A superoperator of
this form projects the total state $\rho$ onto a tensor product
state, i.~e., onto a state without any statistical correlations
between system and environment. Many examples for this
product-state projection are known in the fields of quantum
optics, decoherence, quantum Brownian motion, quantum measurement
theory, and coherent and optimal quantum control. It is typically
applicable in the case of weak system-environment couplings. The
corresponding perturbation expansion is usually restricted to the
second order (known as Born approximation\index{Born
approximation}), from which one derives, with the help of certain
further assumptions, a Markovian quantum master equations in
Lindblad form \cite{GORINI,LINDBLAD,SPOHN}.

A possible approach to large deviations from Markovian behavior
consists in carrying out the perturbation expansion to higher
orders in the system-environment coupling. However, this approach
is often limited by the increasing complexity of the resulting
equations of motion. Moreover, the perturbation expansion may not
converge uniformly in time, such that higher orders only improve
the quality of the approximation of the short-time behavior, but
completely fail in the long-time limit \cite{BBP}.

We will discuss here a further strategy for the treatment of
highly non-Markovian processes which is based on the use of a
correlated projection superoperator
\cite{CPS,BGM,BUDINI06,BUDINI05,GASPARD1,GASPARD2}. By contrast to
the product-state projection, a correlated projection
superoperator projects the total state $\rho$ onto a
system-environment state that contains statistical correlations
between certain system and environment states. We will discuss a
representation theorem for a large class of such projections which
are appropriate for the application of the projection operator
techniques, and develop a corresponding non-Markovian
generalization of the Lindblad equation.

\section{The standard projection operator method}\label{STANDARD-PROJECTION-OPERATORS}
We investigate an open quantum system\index{open quantum system}
$S$ that is coupled to some environment\index{environment} $E$.
The corresponding Hilbert spaces are denoted by ${\mathcal{H}}_S$
and ${\mathcal{H}}_E$, respectively. The state space of the
composite system is thus given by the tensor product\index{tensor
product} space
\begin{equation}
 {\mathcal{H}}={\mathcal{H}}_S\otimes{\mathcal{H}}_E.
\end{equation}
The states of the composite system are represented by density
matrices\index{density matrix} $\rho$ on ${\mathcal{H}}$
satisfying the physical conditions of the positivity and the
normalization:
\begin{equation}
 \rho \geq 0, \qquad {\mathrm{tr}}\rho = 1,
\end{equation}
where ${\mathrm{tr}}$ denotes the trace taken over the total state
space ${\mathcal{H}}$. The partial traces over ${\mathcal{H}}_S$
and ${\mathcal{H}}_E$ will be denoted by ${\mathrm{tr}}_S$ and
${\mathrm{tr}}_E$.

\subsection{Nakajima-Zwanzig projection operator technique}
A central goal of the theory is to develop efficient strategies
for the description of the behavior of the reduced density
matrix\index{reduced density matrix} which is determined by the
partial trace over the environmental state space,
\begin{equation}
 \rho_S = {\mathrm{tr}}_E \rho.
\end{equation}
The basic idea of the projection operator
techniques\index{projection operator techniques} is to regard the
operation of taking the partial trace over $E$ formally as a map
${\mathcal{P}}$ defined by
\begin{equation} \label{STANDARD-PROJECTION}
 {\mathcal{P}}\rho = ({\mathrm{tr}}_E\rho) \otimes \rho_0.
\end{equation}
For a fixed environmental state $\rho_0$ this defines a linear
transformation which maps any density operator $\rho$ on the total
state space ${\mathcal{H}}$ to a density operator
${\mathcal{P}}\rho$ on ${\mathcal{H}}$ and has the property of a
projection operator:
\begin{equation} \label{PROJECTION}
 {\mathcal{P}}^2 = {\mathcal{P}}.
\end{equation}
Being a map acting on operators, ${\mathcal{P}}$ is often called a
projection superoperator. The complementary projection is defined
by
\begin{equation}
 {\mathcal{Q}}=I-{\mathcal{P}},
\end{equation}
$I$ being the identity map. Note that according to
Eq.~(\ref{STANDARD-PROJECTION}) the reduced system's state is
obtained from the projection ${\mathcal{P}}\rho$ by taking the
partial trace over the environment:
\begin{equation}
 \rho_S = {\mathrm{tr}}_E \{ {\mathcal{P}} \rho \}.
\end{equation}

The Hamiltonian of the composite system is of the form
\begin{equation}
 H=H_0+H_I,
\end{equation}
where $H_0$ denotes the unperturbed part, usually given by the sum
of a system Hamiltonian $H_S$ and an environmental Hamiltonian
$H_E$, and $H_I$ represents the interaction. In many cases it is
convenient to formulate the dynamics in the interaction picture
with respect to $H_0$ in which the density matrix $\rho(t)$ of the
total system is governed by the von Neumann equation\index{von
Neumann equation}
\begin{equation}
 \frac{d}{d t}\rho(t) = -i [H_I(t),\rho(t)] \equiv
 {\mathcal{L}}(t)\rho(t).
\end{equation}
The operator
\begin{equation}
 H_I(t) = e^{i H_0t} H_I e^{-i H_0t}
\end{equation}
represents the Hamiltonian in the interaction picture and
${\mathcal{L}}(t)$ the corresponding Liouville
superoperator\index{Liouville superoperator}.

The Nakajima-Zwanzig (NZ) projection operator technique yields a
closed equation of motion for the relevant part
${\mathcal{P}}\rho(t)$ of the density matrix and, hence, for the
reduced density matrix $\rho_S(t)$. To simplify the presentation
we assume that the condition
\begin{equation}
 {\mathcal{P}}{\mathcal{L}}(t_1){\mathcal{L}}(t_2)\ldots
 {\mathcal{L}}(t_{2n+1}){\mathcal{P}} = 0
\end{equation}
holds true. This condition is in fact satisfied in many
applications. Moreover, we suppose that the initial state
satisfies ${\mathcal{P}}\rho(0)=\rho(0)$. The projection
${\mathcal{P}}\rho(t)$ is then governed by a homogeneous
integrodifferential equation, the Nakajima-Zwanzig
equation\index{Nakajima-Zwanzig equation}:
\begin{equation} \label{NZ-GEN}
 \frac{d}{d t} {\mathcal{P}}\rho(t) = \int_0^t d t_1
 {\mathcal{K}}(t,t_1) {\mathcal{P}}\rho(t_1).
\end{equation}
The memory kernel\index{memory kernel} ${\mathcal{K}}(t,t_1)$ is
given by
\begin{equation} \label{NZ-KERNEL}
 {\mathcal{K}}(t,t_1) = {\mathcal P} {\mathcal{L}}(t)\;
 {\mathrm{T}}\exp\left[\int_{t_1}^t d t_2 {\mathcal{Q}}{\mathcal{L}}(t_2) \right]
 {\mathcal Q} {\mathcal{L}}(t_1) {\mathcal P},
\end{equation}
where ${\mathrm{T}}$ denotes the chronological time ordering.

The memory kernel ${\mathcal{K}}(t,t_1)$ is in general a very
complicated superoperator whose determination is in most cases as
complicated as the solution of the full system's dynamics.
Therefore, one usually tries to determine it by a perturbation
expansion in powers of the strength of the system-environment
coupling. The lowest order contribution is given by the
second-order equation of motion:
\begin{equation} \label{NZ2}
 \frac{d}{d t} {\mathcal{P}}\rho(t) = \int_0^t d t_1
 {\mathcal{P}}{\mathcal{L}}(t){\mathcal{L}}(t_1){\mathcal{P}}\rho(t_1).
\end{equation}
Higher orders are obtained with the help of the general expression
(\ref{NZ-KERNEL}) for the memory kernel.

\subsection{Time-convolutionless projection operator technique}\label{TCL-TECHNIQUE}

There exists an alternative expansion technique based on the
projection superoperator ${\mathcal{P}}$ which is known as
time-convolutionless (TCL) projection operator
method\index{time-convolutionless projection operator method}. By
contrast to the NZ approach, the TCL method leads to an equation
of motion for the relevant part of the density matrix which
represents a time-local differential equation of the general form
\begin{equation} \label{TCL-GEN}
 \frac{d}{d t} {\mathcal{P}}\rho(t) =
 {\mathcal{K}}(t) {\mathcal{P}}\rho(t).
\end{equation}
Here, ${\mathcal{K}}(t)$ is a time-dependent superoperator, called
the TCL generator\index{TCL generator}. It should be stressed that
the TCL equation (\ref{TCL-GEN}) describes non-Markovian dynamics,
although it is local in time and does not involve an integration
over the system's past. In fact, the TCL equation takes into
account all memory effects through the explicit time-dependence of
the generator ${\mathcal{K}}(t)$.

To obtain the time-local form of the TCL equation one eliminates
the dependence of the future time evolution on the system's
history through the introduction of the backward propagator into
the Nakajima-Zwanzig equation. This enables one to express the
density matrix at previous times $t_1<t$ in terms of the density
matrix at time $t$ and to derive an exact time-local equation of
motion. We remark that the backward propagator and, hence, also
the TCL generator may not exist, typically at isolated points of
the time axis. This may happen for very strong system-environment
couplings and/or long integration times; an example is discussed
in \cite{TheWork}.

Again, one can develop a systematic perturbation expansion for the
TCL generator which takes the form
${\mathcal{K}}(t)={\mathcal{K}}_2(t)+{\mathcal{K}}_4(t)+\ldots$
The various orders of this expansion can be expressed through the
ordered cumulants\index{ordered cumulants}
\cite{KUBO63,ROYER72,KAMPEN1,KAMPEN2} of the Liouville
superoperator ${\mathcal{L}}(t)$. For instance, the contributions
of second and fourth order to the TCL generator are given by
\cite{TheWork}
\[
 {\mathcal{K}}_2(t) = \int_0^t d t_1
 {\mathcal{P}}{\mathcal{L}}(t){\mathcal{L}}(t_1){\mathcal{P}},
\]
and
\begin{eqnarray*}
 {\mathcal{K}}_4(t) &=& \int_0^td t_1\int_0^{t_1}d t_2\int_0^{t_2}d t_3 \\
 && \times \Big[
 {\mathcal P}{\mathcal{L}}(t){\mathcal{L}}(t_1){\mathcal{L}}(t_2)
 {\mathcal{L}}(t_3){\mathcal P}
 -{\mathcal P}{\mathcal{L}}(t){\mathcal{L}}(t_1){\mathcal P}
  {\mathcal{L}}(t_2){\mathcal{L}}(t_3){\mathcal P} \\
 && \;
 -{\mathcal P}{\mathcal{L}}(t){\mathcal{L}}(t_2){\mathcal P}
  {\mathcal{L}}(t_1){\mathcal{L}}(t_3){\mathcal P}
  -{\mathcal P}{\mathcal{L}}(t){\mathcal{L}}(t_3){\mathcal P}
  {\mathcal{L}}(t_1){\mathcal{L}}(t_2){\mathcal P}
 \Big]. \\
\end{eqnarray*}
In second order the TCL master equation takes the form
\begin{equation} \label{TCL2}
 \frac{d}{d t} {\mathcal{P}}\rho(t) = \int_0^t d t_1
 {\mathcal{P}}{\mathcal{L}}(t){\mathcal{L}}(t_1){\mathcal{P}}\rho(t),
\end{equation}
which should be contrasted to the NZ equation (\ref{NZ2}).

It is important to realize that the NZ and the TCL technique lead
to equations of motion with entirely different structures and
that, therefore, also the mathematical structure of their
solutions are quite different in any given order \cite{ROYER03}.
It is difficult to formulate general conditions that allow to
decide for a given model whether the NZ or the TCL approach is
more efficient. The assessment of the quality of the approximation
obtained generally requires the investigation of higher orders of
the expansion, or else the comparison with numerical simulations
or with certain limiting cases that can be treated analytically.
It turns out that in many cases the degree of accuracy obtained by
both methods are of the same order of magnitude. In these cases
the TCL approach is of course to be preferred because it is
technically much simpler to deal with.

In the NZ equation (\ref{NZ-GEN}) as well as in the TCL equation
(\ref{TCL-GEN}) we made use of the initial condition
${\mathcal{P}}\rho(0)=\rho(0)$. According to the definition
(\ref{STANDARD-PROJECTION}) of the projection ${\mathcal{P}}$ this
condition is equivalent to the assumption that $\rho(0)$
represents an uncorrelated tensor product initial state,
$\rho(0)=\rho_S(0)\otimes\rho_0$. For a correlated initial
state\index{correlated initial state} one has to add a certain
inhomogeneity to the right-hand side of the NZ or the TCL equation
which involves the initial conditions through the complementary
projection ${\mathcal{Q}}\rho(0)=(I-{\mathcal{P}})\rho(0)$. A
general method for the treatment of such correlated initial states
within the TCL technique is described in \cite{TheWork}; for a
recent study on their influence in weakly coupled systems see also
Refs.~\cite{TASAKI1,TASAKI2}.

\subsection{Markovian limit and quantum dynamical semigroups}
With the standard projection defined in
Eq.~(\ref{STANDARD-PROJECTION}), the TCL equation (\ref{TCL2}) is
equivalent to the following master equation for the reduced
density matrix,
\begin{equation} \label{TCL2-MASTER-EQ}
 \frac{d}{d t} \rho_S(t) = -\int_0^t d t_1
 {\mathrm{tr}}_E \{ [H_I(t),[H_I(t_1),\rho_S(t)\otimes\rho_0]] \}.
\end{equation}
This equation provides an appropriate starting point for an
approximation scheme which is known as Born-Markov approximation
and which eventually leads to a Markovian quantum master equation
in Lindblad form\index{Lindblad equation}
\begin{eqnarray} \label{LINDBLAD-EQ}
 \frac{d}{d t} \rho_S(t) &=& {\mathcal{K}}\rho_S(t) \nonumber \\
 &=& -i \left[ H_S,\rho_S(t) \right]
 + \sum_{\lambda} \left(
 R_{\lambda} \rho_S(t) R^{\dagger}_{\lambda}
 - \frac{1}{2} \left\{
 R^{\dagger}_{\lambda} R_{\lambda}, \rho_S(t)\right\} \right).
\end{eqnarray}
Here, ${\mathcal{K}}$ is a time-independent generator, the
Lindblad generator, involving a Hermitian operator $H_S$ and
arbitrary system operators $R_{\lambda}$. Therefore, it generates
state transformations of the form
\begin{equation} \label{QDM}
 \Phi_t: \rho_S(0) \mapsto \rho_S(t), \qquad
 \Phi_t = e^{\mathcal{K}t}.
\end{equation}
$\Phi_t$ is called a quantum dynamical map\index{quantum dynamical
map} and the set of transformations
\[
 \{\Phi_t|t\geq 0\}
\]
is referred to as quantum dynamical semigroup\index{quantum
dynamical semigroup}. Under certain technical conditions, it can
be shown that the form of the Lindblad generator guarantees the
preservation of the positivity and normalization of the density
matrix, as well as the complete positivity\index{completely
positive maps} of the dynamical transformation $\Phi_t$. Vice
versa, any completely positive quantum dynamical semigroup has a
generator of the form (\ref{LINDBLAD-EQ}). This is the well-known
Gorini-Kossakowski-Sudarshan-Lindblad theorem
\cite{GORINI,LINDBLAD}.

The microscopic derivation of the master equation
(\ref{LINDBLAD-EQ}) from the TCL equation (\ref{TCL2-MASTER-EQ})
requires the validity of several approximations, the most
important one being the so-called Markov
approximation\index{Markov approximation}. This approximation
presupposes a rapid decay of the two-point correlation
functions\index{correlation functions} of those environmental
operators that describe the system-environment coupling. More
precisely, if $\tau_E$ describes the temporal width of these
correlations and $\tau_R$ the relaxation time\index{relaxation
time} of the system, the Markov approximation demands that
\begin{equation} \label{MARKOV-COND}
 \tau_E \ll \tau_R.
\end{equation}
This means that the environmental correlation
time\index{correlation time} $\tau_E$ is short compared to the
open system's relaxation time $\tau_R$.

The Markov approximation is justified in many cases of physical
interest. Examples of application are the quantum optical master
equation describing the interaction of radiation with matter, and
the master equation for a test particle in a quantum gas
\cite{VACCHINI1,VACCHINI2,HORNBERGER}. However, strong couplings
or interactions with low-temperature reservoirs can lead to large
correlations resulting in long memory times and in a failure of
the Markov approximation. In the following, the quantum dynamics
of an open system is said to be non-Markovian\index{non-Markovian
quantum dynamics} if the time-evolution of its reduced density
matrix cannot be described (to the desired degree of accuracy) by
means of a closed master equation with a (possibly time-dependent)
generator in Lindblad form.

If the two-point environmental correlation functions do not decay
rapidly in time the second order of the expansion cannot, in
general, be expected to give an accurate description of the
dynamics. For instance, this situation arises for the spin star
model discussed in Ref.~\cite{BBP}, where the second-order
generator of the master equation increases linearly with time such
that the Born-Markov approximation simply does not exist.

More importantly, the standard Markov condition\index{Markov
condition} (\ref{MARKOV-COND}) alone does {\textit{not}}
guarantee, in general, that the Markovian master equation provides
a reasonable description of the dynamics. This situation can occur
for finite and/or structured reservoirs that cannot be represented
by a Bosonic field or a collection of harmonic oscillator modes.
In such cases a detailed investigation of the influence of
higher-order correlations is indispensable in order to judge the
quality of a given order. The model discussed in Ref.~\cite{BGM}
represents an example for which the standard Markov condition
{\textit{is}} satisfied although the expansion based on the
projection (\ref{STANDARD-PROJECTION}) completely fails if one
truncates the expansion at any finite order. In such cases strong
non-Markovian dynamics is induced through the behavior of
higher-order correlation functions.

We conclude that in general one can judge the quality of a given
projection superoperator and a given expansion technique that is
based on it only by an investigation of the structure of higher
orders. The standard projection and the corresponding Lindblad
equation are not reliable if higher orders lead to contributions
that are not bounded in time, signifying the non-uniform
convergence of the perturbation expansion \cite{BGM}.

\section{Correlated projection superoperators}\label{Sec-CPS}
\index{correlated projection superoperators}

The performance of the projection operator techniques depends of
course on the properties of the microscopic model under study, in
particular on the structure of the correlation functions of the
model. However, it also depends strongly on the choice of the
superoperator ${\mathcal{P}}$. Several extensions of the standard
projection (\ref{STANDARD-PROJECTION}) and modifications of the
expansion technique have been proposed in the literature (see,
e.g., Refs.~\cite{OPPENHEIM1,OPPENHEIM2,FRIGERIO}).

The projection defined by Eq.~(\ref{STANDARD-PROJECTION}) projects
any state $\rho$ onto a tensor product $\rho_S \otimes \rho_0$
that describes a state without statistical correlations between
the system and its environment. Here, we introduce a more general
class of projection superoperators that project onto correlated
system-environment states and are therefore able to describe
strong correlations and non-Markovian effects \cite{CPS}.

\subsection{General conditions}\label{GEN-COND}
We assume that our new class of maps ${\mathcal{P}}$ represent
superoperators with the property of a projection, i.e.,
${\mathcal{P}}^2={\mathcal{P}}$. As a consequence, the whole
machinery of the projection operator techniques described in
Sec.~\ref{STANDARD-PROJECTION-OPERATORS} can be applied also to
the new class of correlated maps.

Within the projection operator techniques the projection
${\mathcal{P}}\rho$ should represent a suitable approximation of
$\rho$. We therefore require that for any physical state $\rho$
the projection ${\mathcal{P}}\rho$ is again a physical state,
i.~e., a positive operator with unit trace. This means that
${\mathcal{P}}$ is a positive and trace preserving
map\index{positive maps}, namely
\begin{equation}
 \rho \geq 0 \Longrightarrow {\mathcal{P}}\rho \geq 0,
 \qquad
 {\mathrm{tr}}\{{\mathcal{P}}\rho\}={\mathrm{tr}}\rho.
\end{equation}

Our class of projection operators is assumed to consist of maps of
the following general form,
\begin{equation} \label{ITIMESLAMBDA}
 {\mathcal{P}} = I_S \otimes \Lambda.
\end{equation}
Here, $I_S$ denotes the unit map acting on operators on
${\mathcal{H}}_S$, and $\Lambda$ is a linear map that transforms
operators on ${\mathcal{H}}_E$ into operators on
${\mathcal{H}}_E$. A projection superoperator of this form leaves
the system $S$ unchanged and acts nontrivially only on the
variables of the environment $E$. As a consequence of the
positivity of ${\mathcal{P}}$ and of condition
(\ref{ITIMESLAMBDA}) the map $\Lambda$ must be $N_S$-positive,
where $N_S$ is the dimension of ${\mathcal{H}}_S$. In the
following we use the stronger condition that $\Lambda$ is
completely positive\index{completely positive maps}, because
completely positive maps allow for a simple mathematical
characterization (see Sec.~\ref{SEC-REPRESENTATION}).

Let us discuss the physical implications of these conditions.
According to Eqs.~(\ref{PROJECTION}) and (\ref{ITIMESLAMBDA}) the
map $\Lambda$ must itself be a projection, namely
$\Lambda^2=\Lambda$. Moreover, since ${\mathcal{P}}$ is
trace-preserving, the map $\Lambda$ must also be trace-preserving.
Hence, we find that $\Lambda$ represents a completely positive and
trace-preserving map (CPT map\index{CPT maps}, or quantum
channel\index{quantum channel}) which operates on the variables of
the environment and has the property of a projection. The action
of ${\mathcal{P}}$ may also be interpreted as that of a
generalized quantum measurement\index{generalized quantum
measurement} which is carried out on the environment. A further
physically reasonable consequence of the positivity of $\Lambda$
and of Eq.~(\ref{ITIMESLAMBDA}) is that ${\mathcal{P}}$ maps
product states\index{product state} to product states, and, more
generally, separable (classically correlated) states to separable
states\index{separable state}. This means that the application of
${\mathcal{P}}$ does not create entanglement\index{entangled
state} between the system and its environment.

Using Eq.~(\ref{ITIMESLAMBDA}) and the fact that $\Lambda$ is
trace-preserving we get
\begin{equation} \label{RHO-S-P}
 \rho_S \equiv {\mathrm{tr}}_E \rho = {\mathrm{tr}}_E \{
 {\mathcal{P}} \rho \}.
\end{equation}
This relation connects the density matrix of the reduced system
with the projection of a given state $\rho$ of the total system.
It states that, in order to determine $\rho_S$, we do not really
need the full density matrix $\rho$, but only its projection
${\mathcal{P}}\rho$. Thus, ${\mathcal{P}}\rho$ contains the full
information needed to reconstruct the reduced system's state.

\subsection{Representation theorem}\label{SEC-REPRESENTATION}
\index{representation theorem}

What is the explicit structure of the projection superoperators
satisfying the basic conditions formulated above? This question is
answered by a representation theorem \cite{CPS} which states that
${\mathcal{P}}$ fulfills the condition of Sec.~\ref{GEN-COND} if
and only if it can be written in the form
\begin{equation} \label{PROJECTION-GENFORM}
 {\mathcal{P}}\rho = \sum_i {\mathrm{tr}}_E \{ A_i \rho \}
 \otimes B_i,
\end{equation}
where $\{A_i\}$ and $\{B_i\}$ are two sets of linear independent
Hermitian operators satisfying the relations
\begin{eqnarray}
 {\mathrm{tr}}_E \{ B_i A_j \} &=& \delta_{ij}, \label{BjAi} \\
 \sum_i ({\mathrm{tr}}_E B_i) A_i &=& I_E, \label{TRACE-PRESERVING} \\
 \sum_i A_i^T \otimes B_i &\geq& 0. \label{COND-POS}
\end{eqnarray}
Equation (\ref{BjAi}) guarantees that ${\mathcal{P}}$ is a
projection superoperator, Eq.~(\ref{TRACE-PRESERVING}) ensures
that ${\mathcal{P}}$ is trace-preserving, and Eq.~(\ref{COND-POS})
is equivalent to the condition of complete
positivity\index{completely positive maps} ($T$ denotes the
transposition\index{transposition}).

The standard projection (\ref{STANDARD-PROJECTION}) that projects
onto an uncorrelated tensor product state is obviously of the form
of Eq.~(\ref{PROJECTION-GENFORM}). In fact, if we take a single
$A=I_E$ and a single $B=\rho_0$ the conditions (\ref{BjAi}),
(\ref{TRACE-PRESERVING}), and (\ref{COND-POS}) are trivially
satisfied and Eq.~(\ref{PROJECTION-GENFORM}) obviously reduces to
Eq.~(\ref{STANDARD-PROJECTION}). Of course, a projection
${\mathcal{P}}\rho$ of the form of Eq.~(\ref{PROJECTION-GENFORM})
does not in general represent a simple product state. We therefore
call such ${\mathcal{P}}$ correlated projection superoperators.
They project onto states that contain statistical correlations
between the system $S$ and its environment $E$. In the following
we will consider the case that one can find a representation of
the projection with positive operators
\begin{equation}
 A_i \geq 0, \qquad B_i \geq 0.
\end{equation}
Equation (\ref{COND-POS}) is then trivially satisfied. Without
restriction we may assume that the $B_i$ are normalized to unit
trace,
\begin{equation}
 {\mathrm{tr}}_E B_i = 1,
\end{equation}
such that condition (\ref{TRACE-PRESERVING}) reduces to the simple
form
\begin{equation}
 \sum_i A_i = I_E.
\end{equation}
Under these conditions ${\mathcal{P}}$ projects any state $\rho$
onto a state which can be written as a sum of tensor products of
positive operators. In the theory of entanglement (see, e.g., the
recent review \cite{HORODECKI}) such states are called
separable\index{separable state} or classically correlated. Using
a projection superoperator of this form, one thus tries to
approximate the total system's states by a classically correlated
state. The general representation of
Eq.~(\ref{PROJECTION-GENFORM}) includes the case of projection
superoperators that project onto inseparable, entangled quantum
states\index{entangled state}. We will not pursue here this
possibility further, and restrict ourselves to positive $A_i$ and
$B_i$ in the following.

\subsection{Correlated initial states}\label{CORR-INIT}
\index{correlated initial state}

As mentioned already in Sec.~\ref{TCL-TECHNIQUE} a homogeneous NZ
or TCL equation of motion presupposes a tensor product initial
state if one uses the standard projection superoperator
(\ref{STANDARD-PROJECTION}). However, this is no longer true for
the correlated projection defined by
Eq.~(\ref{PROJECTION-GENFORM}). In fact, the general condition for
the absence of an inhomogeneous term in the NZ equation
(\ref{NZ-GEN}) or the TCL equation (\ref{TCL-GEN}) is given by
\begin{equation}
 {\mathcal{P}}\rho(0)=\rho(0).
\end{equation}
According to Eq.~(\ref{PROJECTION-GENFORM}) this condition is
equivalent to the assumption that $\rho(0)$ takes the form
\begin{equation} \label{INIT}
 \rho(0) = \sum_i \rho_i(0) \otimes B_i,
\end{equation}
where
\begin{equation}
 \rho_i(0)={\mathrm{tr}}_E\{A_i\rho(0)\} \geq 0.
\end{equation}
Equation (\ref{INIT}) represents in general a correlated initial
state. Hence, a great advantage of the correlated projection
superoperators is given by the fact that they allows the treatment
of correlated initial states by means of a homogeneous NZ or TCL
equation \cite{CPS}.

\subsection{Conservation laws}\label{CONS}
\index{conservation laws}

A crucial step in applications of the correlated projection
operator technique is the construction of an appropriate
projection superoperator ${\mathcal{P}}$. An important strategy
for this construction is to take into account the known conserved
quantities of the model under study.

Suppose $C$ is a conserved observable. A good choice for the
projection superoperator ${\mathcal{P}}$ will then be a projection
that leaves invariant the expectation value of $C$, i.e. that
satisfies the relation
\begin{equation} \label{C-COND-1}
 {\mathrm{tr}} \{ C\rho \}
 = {\mathrm{tr}} \{ C ({\mathcal{P}} \rho) \}.
\end{equation}
To bring this relation into a more convenient form we introduce
the adjoint ${\mathcal{P}}^{\dagger}$ of the projection
superoperator ${\mathcal{P}}$. The adjoint map\index{adjoint map}
is defined with the help of the Hilbert-Schmidt scalar
product\index{Hilbert-Schmidt scalar product}
\begin{equation}
 (X,Y) = {\mathrm{tr}}\{X^{\dagger}Y\}
\end{equation}
for the space of operators acting on the state space of the total
system through the relation
\[
 (X,{\mathcal{P}}Y)=({\mathcal{P}}^{\dagger}X,Y).
\]
This allows us to write Eq.~(\ref{C-COND-1}) in the form
${\mathrm{tr}}\{ C\rho \} = {\mathrm{tr}} \{
({\mathcal{P}}^{\dagger}C) \rho \}$. Requiring this to hold for
all states $\rho$ we get the relation
\begin{equation} \label{C-COND-2}
 {\mathcal{P}}^{\dagger} C = C.
\end{equation}

The adjoint of the projection (\ref{PROJECTION-GENFORM}) is
obtained by interchanging the role of $A_i$ and $B_i$. Hence,
condition (\ref{C-COND-2}) can be written explicitly as
\begin{equation} \label{C-COND-3}
 {\mathcal{P}}^{\dagger} C = \sum_i {\mathrm{tr}}_E \{ B_i C \}
 \otimes A_i = C.
\end{equation}
This equation represents a condition for the projection
superoperator ${\mathcal{P}}$ on the basis of a known conserved
quantity of the underlying model. It ensures that the projection
superoperator leaves invariant this quantity and that the
effective description respects the corresponding conservation law.

\section{Generalization of the Lindblad equation}\label{DYNAMICS}
\index{Lindblad equation!non-Markovian generalization}

Once a projection superoperator has been chosen the projection
${\mathcal{P}}\rho(t)$ of the time-dependent total system's state
$\rho(t)$ is, according to Eq.~(\ref{PROJECTION-GENFORM}),
uniquely determined by the dynamical variables
\begin{equation}
 \rho_i(t) = {\mathrm{tr}}_E\{ A_i \rho(t) \}.
\end{equation}
To be specific we assume in the following that that the index $i$
takes on the values $i=1,2,\ldots,n$. Since we require that the
$A_i$ are positive, the $\rho_i(t)$ are positive operators. From
Eq.~(\ref{RHO-S-P}) we find the connection to the reduced density
matrix,
\begin{equation} \label{REDUCED-DENSITY}
 \rho_S(t) = \sum_i \rho_i(t),
\end{equation}
and the normalization condition takes the form
\begin{equation}
 {\mathrm{tr}}_S \rho_S(t) = \sum_i {\mathrm{tr}}_S \rho_i(t) = 1.
\end{equation}
Hence, we see that the reduced system's state is uniquely
determined by a set of $n$ (unnormalized) density operators
$\rho_i(t)$.

Our formulation leads to a natural question, namely what is the
analog of the Lindblad equation (\ref{LINDBLAD-EQ}) in the case of
a correlated projection superoperator? To answer this question we
first observe that the time-evolution leads to a transformation of
the form
\begin{equation}
 \{ \rho_i(0) \} \mapsto \{ \rho_i(t) \},
\end{equation}
transforming any initial set of positive operators $\rho_i(0)\geq
0$ into another set of positive operators $\rho_i(t)\geq 0$ at
time $t > 0$. This transformation can conveniently be described
with the help of an auxiliary $n$-dimensional Hilbert space
$\mathbf{C}^n$ and a fixed orthonormal basis $\{ |i\rangle \}$ for
this space. Then we can identify the collection of densities
$\rho_i(t)$ with a density matrix $\varrho(t)$ on the extended
space\index{extended state space}
\begin{equation}
 {\mathcal{H}}_{\mathrm{ext}} = {\mathcal{H}}_S \otimes \mathbf{C}^n
\end{equation}
through the relation
\begin{equation} \label{EXTENDED-SPACE}
 \varrho(t) = \sum_i \rho_i(t) \otimes |i\rangle\langle i|.
\end{equation}
This density matrix can be regarded as a block diagonal matrix
\begin{equation}
 \varrho(t) =
  \left( \begin{array}{cccc}
  \rho_1(t) & 0 & \cdots &  0          \\
  0            & \rho_2(t) & \cdots & 0 \\
  0            & 0                   & \ddots    & 0 \\
  0            & 0                   & \cdots    & \rho_n(t)
                                     \end{array}    \right)
\end{equation}
with blocks $\rho_i(t)$ along the main diagonal. Moreover, the
reduced density matrix $\rho_S(t)$ is obtained by the partial
trace of $\varrho(t)$ taken over the auxiliary space.

In close analog to Eq.~(\ref{QDM}) the dynamics may now be viewed
as a transformation
\begin{equation}
 V_t: \varrho(0) \mapsto \varrho(t)
\end{equation}
that preserves the block diagonal structure. It is important to
emphasize that $V_t$ is {\textit{not}} a quantum dynamical
map\index{quantum dynamical map} in the usual sense because it is
{\textit{not}} an operation on the space of states of the reduced
system, but rather a map on the extended state space. In fact, the
transition from $\varrho(0)$ to the reduced density matrix
$\rho_S(0)=\sum_i \rho_i(0)$ is connected with a loss of
information on the initial correlations, such that from the mere
knowledge of $\rho_S(0)$ the dynamical behavior cannot be
reconstructed.

It may be shown that $V_t$ can be extended to a completely
positive map\index{completely positive maps} for operators on
${\mathcal{H}}_{\mathrm{ext}}$. Hence, we can construct an
embedding\index{embedding into a Lindblad dynamics} of the
dynamical transformation into a Lindblad dynamics on the extended
state space. This is achieved by the requirement that there exist
a Lindblad generator ${\mathcal{K}}$ acting on operators of the
extended state space which preserves the block diagonal structure:
\begin{equation} \label{L-PROP}
 {\mathcal{K}} \left( \sum_i \rho_i \otimes |i\rangle\langle i|
 \right) = \sum_i {\mathcal{K}}_i(\rho_1,\ldots,\rho_n)
 \otimes |i\rangle\langle i|,
\end{equation}
such that the time-evolution can be represented in the form
\begin{equation} \label{EMBEDDING}
 \sum_i \rho_i(t) \otimes |i\rangle\langle i|
 = e^{{\mathcal{K}}t} \left(
 \sum_i \rho_i(0) \otimes |i\rangle\langle i| \right).
\end{equation}
One can show that a Lindblad generator ${\mathcal{K}}$ with this
property exists if and only if the densities $\rho_i(t)$ obey the
master equation
\begin{equation} \label{GEN-MASTER}
 \frac{d}{d t} \rho_i(t) = -i \left[ H^i,\rho_i(t) \right]
 + \sum_{j \lambda} \left(
 R^{ij}_{\lambda} \rho_j(t) R^{ij\dagger}_{\lambda}
 - \frac{1}{2} \left\{
 R^{ji\dagger}_{\lambda} R^{ji}_{\lambda}, \rho_i(t) \right\} \right).
\end{equation}
The $H^i$ are Hermitian operators on ${\mathcal{H}}_S$, while the
$R^{ij}_{\lambda}$ may be arbitrary operators on
${\mathcal{H}}_S$. The details of the proof of this statement can
be found in Ref.~\cite{CPS}.

Equation (\ref{GEN-MASTER}) represents the desired non-Markovian
generalization of the Lindblad equation\index{Lindblad
equation!non-Markovian generalization} (\ref{LINDBLAD-EQ}) for the
case of a classically correlated projection superoperator. This
equation has many physical applications. In fact, master equations
of the form of Eq.~(\ref{GEN-MASTER}) have been derived by several
authors and applied to various models featuring pronounced
non-Markovian effects
\cite{CPS,BGM,BUDINI06,BUDINI05,GASPARD1,GASPARD2}.

\section{Conclusions}\label{CONCLU}
We have reviewed the theoretical treatment of non-Markovian
quantum dynamics within the framework of the projection operator
techniques. It has been shown that an efficient description of
strong non-Markovian effects is made possible through the
construction of correlated projection superoperators
${\mathcal{P}}$. The central idea behind this construction is to
take into account large system-environment correlations by an
extension of the set of dynamical variables. In fact, employing a
correlated projection superoperator instead of a product-state
projection, one enlarges the set of dynamical variables from the
reduced density matrix $\rho_S$ to a collection of densities
$\rho_i$ describing system states that are correlated with certain
environmental states.

General physical conditions for a large class of correlated
projection superoperators have been formulated, demanding
essentially that ${\mathcal{P}}$ can be expressed in terms of a
projective quantum channel that operates on the environmental
variables. These conditions lead to a representation theorem for
correlated projection superoperators and to a non-Markovian
generalization of the Lindblad equation that is capable of
modelling long memory times and large initial correlations, while
preserving the physical conditions of positivity and
normalization.

The method developed here has many applications to physically
relevant models featuring non-Markovian dynamics. The investigated
class of projections does not exhaust all possibilities. Future
investigations should include the formulation of further classes
of correlated projections, the study of time-dependent generators,
as well as the application of correlated maps that project onto
nonseparable, entangled quantum states.

\subsection*{Acknowledgement}
I would like to thank Francesco Petruccione, Daniel Burgarth, Jan
Fischer, Jochen Gemmer, Mathias Michel, and Bassano Vacchini for
many fruitful discussions.


\begin{thebibliography}{99}
\bibitem{TheWork} H. P. Breuer, F. Petruccione:
                  \textit{The Theory of Open Quantum Systems}
                  (Oxford University Press, Oxford 2007)
\bibitem{REIBOLD} F. Haake, R. Reibold: Phys. Rev. A
                  \textbf{32}, 2462 (1985)
\bibitem{HU} B. L. Hu, J. P. Paz, Y. Zhang:
             Phys. Rev. D \textbf{45}, 2843 (1992)
\bibitem{INGOLD} H. Grabert, P. Schramm, G.-L. Ingold:
                 Phys. Rep. \textbf{168}, 115 (1988)
\bibitem{BUZEK} P. \v{S}telmachovi\v{c}, V. Bu\v{z}ek:
                Phys. Rev. A \textbf{64}, 062106 (2001);
                Phys. Rev. A \textbf{67}, 029902(E) (2003)
\bibitem{LOSS} J. Schliemann, A. Khaetskii, D. Loss: J. Phys.:
               Condens. Matter \textbf{15}, R1809 (2003)
\bibitem{GEMMER2005} J. Gemmer, M. Michel: Europhys. Lett. \textbf{73}, 1 (2006)
\bibitem{GMM} J. Gemmer, M. Michel, G. Mahler, \textit{Quantum Thermodynamics},
              Lecture Notes in Physics, vol 657 (Springer, Berlin Heidelberg New York 2004)
\bibitem{FOURIER} M. Michel, G. Mahler, J. Gemmer: Phys. Rev.
                  Lett. \textbf{95}, 180602 (2005)
\bibitem{NAKAJIMA} S. Nakajima: Progr. Theor. Phys. \textbf{20}, 948 (1958)
\bibitem{ZWANZIG} R. Zwanzig: J. Chem. Phys. \textbf{33}, 1338 (1960)
\bibitem{HAAKE} F. Haake: \textit{Statistical Treatment of Open Systems},
                Springer Tracts in Modern Physics, vol 66 (Springer, Berlin 1973)
\bibitem{KUBO} R. Kubo, M. Toda, N. Hashitsume:
               \textit{Statistical Physics II. Nonequilibrium Statistical Mechanics}
               (Springer, Berlin 1991)
\bibitem{SHIBATA77} F. Shibata, Y. Takahashi, N. Hashitsume:
                    J. Stat. Phys. \textbf{17}, 171 (1977)
\bibitem{SHIBATA79} S. Chaturvedi, F. Shibata:
                    Z. Phys. B \textbf{35}, 297 (1979)
\bibitem{SHIBATA80} F. Shibata, T. Arimitsu:
                    J. Phys. Soc. Jap. \textbf{49}, 891 (1980)
\bibitem{SHIBATA99} C. Uchiyama, F. Shibata:
                    Phys. Rev. E \textbf{60}, 2636 (1999)
\bibitem{ROYER2003} A. Royer: Phys. Lett. A \textbf{315}, 335 (2003)
\bibitem{BREUER2004} H. P. Breuer: Phys. Rev. A \textbf{70}, 012106 (2004)
\bibitem{GORINI} V. Gorini, A. Kossakowski, E. C. G. Sudarshan:
                 J. Math. Phys. \textbf{17}, 821 (1976)
\bibitem{LINDBLAD} G. Lindblad: Commun. Math. Phys. \textbf{48}, 119 (1976)
\bibitem{SPOHN} H. Spohn: Rev. Mod. Phys. \textbf{52}, 569 (1980)
\bibitem{BBP} H. P. Breuer, D. Burgarth, F. Petruccione:
              Phys. Rev. B \textbf{70}, 045323 (2004)
\bibitem{CPS} H. P. Breuer: Phys. Rev. A \textbf{75}, 022103 (2007)
\bibitem{BGM} H. P. Breuer, J. Gemmer, M. Michel,
              Phys. Rev. E \textbf{73}, 016139 (2006)
\bibitem{BUDINI06} A. A. Budini: Phys. Rev. A \textbf{74}, 053815 (2006)
\bibitem{BUDINI05} A. A. Budini: Phys. Rev. E \textbf{72}, 056106 (2005)
\bibitem{GASPARD1} M. Esposito, P. Gaspard:
                   Phys. Rev. E \textbf{68}, 066112 (2003)
\bibitem{GASPARD2} M. Esposito, P. Gaspard,
                   Phys. Rev. E \textbf{68}, 066113 (2003)
\bibitem{KUBO63} R. Kubo: J. Math. Phys. \textbf{4}, 174 (1963)
\bibitem{ROYER72} A. Royer: Phys. Rev. A \textbf{6}, 1741 (1972)
\bibitem{KAMPEN1} N. G. van Kampen: Physica \textbf{74}, 215 (1974)
\bibitem{KAMPEN2} N. G. van Kampen: Physica \textbf{74}, 239 (1974)
\bibitem{ROYER03} A. Royer: Aspects of Open Quantum Dynamics.
                  In: \textit{Irreversible Quantum Dynamics}, Lecture Notes in Physics,
                  vol 622, ed by F. Benatti, R. Floreanini (Springer, Berlin Heidelberg New York 2003)
                  pp 47-63
\bibitem{TASAKI1} S. Tasaki et al: Annals of Physics \textbf{322}, 631 (2007)
\bibitem{TASAKI2} K. Yuasa et al: Annals of Physics \textbf{322}, 657 (2007)
\bibitem{VACCHINI1} B. Vacchini: Phys. Rev. Lett. \textbf{84}, 1374 (2000)
\bibitem{VACCHINI2} B. Vacchini: J. Math. Phys. \textbf{42}, 4291 (2001)
\bibitem{HORNBERGER} K. Hornberger: Phys. Rev. Lett. \textbf{97}, 060601 (2006)
\bibitem{OPPENHEIM1} V. Romero-Rochin, I. Oppenheim: Physica A \textbf{155}, 52 (1989)
\bibitem{OPPENHEIM2} V. Romero-Rochin, A. Orsky, I. Oppenheim:
                     Physica A \textbf{156}, 244 (1989)
\bibitem{FRIGERIO} V. Gorini, M. Verri, A. Frigerio:
                   Physica A \textbf{161}, 357 (1989)
\bibitem{HORODECKI} R. Horodecki, P. Horodecki, M. Horodecki, K. Horodecki:
                    \textit{Quantum entanglement}, arXiv:quant-ph/0702225
\end{thebibliography}
\end{document}